\documentclass[aps,
		prl,
		reprint,
        superscriptaddress,
        shortbibliography,
		  nofootinbib,
		floatfix,
		notitlepage
		]{revtex4-1}
\pdfoutput=1

\usepackage{tikz-feynman}

\usepackage{amsmath}
\usepackage{amssymb}
\usepackage{slashed}
\usepackage{color}
\usepackage{hyperref}

\newcommand{\bra}[1]{\langle\,#1\,|}          
\newcommand{\ket}[1]{|\,#1\,\rangle}          %
\newcommand{\ud}{{\mathrm{d}}}
\hfuzz=\maxdimen \tolerance=10000
\vfuzz=\maxdimen \tolerance=10000
\newcommand{\LCm}{{\scriptscriptstyle -}}
\newcommand{\LCp}{{\scriptscriptstyle +}}
\newcommand{\LCpm}{{\scriptscriptstyle \pm}}

\newcommand{\LCperp}{{\scriptscriptstyle \perp}}
\newcommand{\be}{\begin{equation}}
\newcommand{\ee}{\end{equation}}
\newcommand{\bi}{\begin{enumerate}}
\newcommand{\ei}{\end{enumerate}}

\newcommand{\sfp}{{\mathsf{p}}}

\newcommand{\e}{{\mathrm{e}}}

\begin{document}

\title{Scattering with total depletion in strong, focussed fields}

\author{Tim Adamo}
\email{t.adamo@ed.ac.uk}
\affiliation{School of Mathematics \& Maxwell Institute for Mathematical Sciences, University of Edinburgh, EH9 3FD, UK}

\author{Anton Ilderton}
\email{anton.ilderton@ed.ac.uk}
\affiliation{Higgs Centre, School of Physics and Astronomy, University of Edinburgh, EH9 3FD, UK}

\begin{abstract}
Theoretical approaches to QED scattering in strong fields typically treat the field as a fixed background with simple spacetime dependence, such as a plane wave.
Two major challenges are therefore the inclusion of backreaction (e.g.~depletion of the field) and spatial geometry (e.g.~focussing).
We show here that a solution to one problem can solve the other:
even if particle wavefunctions in a chosen focussed background are not known, we show they can be constructed once depletion is accounted for.
We demonstrate this by giving the exact wavefunctions in a flying focus beam {for which all energy is absorbed by particles scattering on it}.
{In addition we use the wavefunctions to obtain a simple expression for the nonlinear Compton scattering amplitude},
comparing with the plane wave case.
Our methods thus open a new avenue of investigation in which two previously challenging effects are simultaneously brought under analytic control.
\end{abstract}
\maketitle

Ultra-intense lasers now access the relativistic, quantum regime of light-matter interactions~\cite{Gonoskov:2021hwf,Fedotov:2022ely}, in which the effective coupling of matter to the laser easily exceeds unity.
This strong field regime demands non-perturbative methods, but typically restricts theory investigations to simple laser models, in particular background (i.e.,~fixed) plane waves.
Two challenging problems are, accordingly, to account analytically for focussing effects~\cite{DiPiazza:2013vra,Heinzl:2017zsr}, and for backreaction on the laser~\cite{Seipt:2016fyu}.
Focussing (i.e.,~spatial field inhomogeneity) opens the door to richer scattering kinematics~\cite{Fedotov2009exact,Gies:2013yxa,Gonoskov:2013ada,DiPiazza:2013vra,Heinzl:2017zsr,Karbstein:2019dxo,DiPiazza:2020wxp,Heinzl:2024cia}, but turns the essential problem of constructing
the wavefunctions which describe particles in the field from a first order ODE in the plane wave case to a typically intractable PDE.
Depletion is conceptually more challenging, as the background field formalism assumes no depletion from the outset, by definition.
Accounting for both focussing and beam depletion analytically is a significant challenge.

However, in this letter we show that a solution to one of these problems can actually solve  \emph{both}: including depletion from the outset can drastically simplify the inclusion of focussing effects. To show this, we construct exact wavefunctions for particles in a {totally} depleting `flying focus' beam, despite the fact that the wavefunctions in the undepleted background (the analogue of Volkov solutions in plane waves~\cite{Wolkow:1935zz}) \emph{are not known}.

{The focal spot of a flying focus laser can be tailored to move at any speed, parallel or anti-parallel to the beam; see~\cite{Froula2018,Jolly:20,FFexpt1,FFexpt2} for experimental realisations. As \emph{focussed} solutions of Maxwell's equations in vacuum, examples date back to~\cite{Brittingham83,Sezginer1985,Hillion92}. Models where the focus counter-propagates at the speed of light, which we consider here, have been used to discuss radiation formation time~\cite{PhysRevA.103.012215}, radiation reaction~\cite{Formanek:2021bpw}, and vacuum birefringence~\cite{Formanek:2023mkx}.}

{Experimental predictions are not the goal of this paper. However, to}  {illustrate} that our methods provide insights into the physics of focussing and depletion, we use our wavefunctions to calculate the amplitude for photon emission from a charged particle in a depleting flying focus beam, and analyse the allowed harmonic ranges of the emitted photons, comparing with the plane wave case.
{While complete depletion of a beam by a single electron is not experimentally relevant, this is a useful toy calculation which demonstrates} that our depleting flying focus {model (below)} is a background \emph{beyond} plane waves where exact analytic amplitude computations are possible.

\paragraph{From depletion to focussing.} {Consider the QED scattering amplitude}
\be\label{SFI0}
    S_{fi} = \bra{\text{out};\beta}S\ket{\alpha;\text{in}} \;, 
\ee
where $S$ is the S-matrix, `in' and `out' label initial and final collections of particles, while $\alpha$ and $\beta$ label \emph{coherent} states of photons. {These model initial and final-state classical fields, given by the expectation value of the gauge field in the coherent state}. {In QED this is used to model (intense) laser fields~\cite{Fedotov:2022ely}. Outside QED, coherent states have recently been applied to the study of gravitational waveforms~\cite{Cristofoli:2021vyo,Cristofoli:2021jas}, and double copy~\cite{Monteiro:2020plf}.}

It is known that $S_{fi}$ in (\ref{SFI0}) is equal (up to normalisation) to the amplitude $\bra{\text{out}}S[A]\ket{\text{in}}$ where $S[A]$ is the S-matrix in the presence of a background $A_\mu$~\cite{Kibble:1965zza,Frantz:1965}. This $A_\mu$, a solution of Maxwell's equations in vacuum, is defined in terms of the coherent states by
\be\label{Fourier-beam}
    A_\mu(x) = \int_\ell \varepsilon^s_\mu (\ell)\alpha_s(\ell)\, \e^{-i\ell\cdot x} + {\bar{\varepsilon}}^s_\mu (\ell){\bar \beta}_s(\ell)\, \e^{+i\ell\cdot x}\,,
\ee
where the $\varepsilon^s_\mu(\ell)$ are polarisation vectors in a chosen gauge and the Fourier coefficients $\alpha_s(\ell)$ and $\beta_s(\ell)$ give mode occupancies for, respectively, incoming and outgoing photons, helicity $s=\pm 1$ and momentum~$\ell_\mu$.   {Taking $\alpha_s=\beta_s$ then means assuming the presence of a field which is unchanged by scattering -- there is no back-reaction on it. It follows that some aspects of back-reaction can be captured by taking $\beta_s\not=\alpha_s$, i.e.~allowing changes to the coherent state. This idea has been used to model energy absorption from classical fields (depletion), both in QED~\cite{Ilderton:2017xbj} and in gravity~\cite{Endlich:2016jgc,Aoude:2023fdm}.} Note that the corresponding background (\ref{Fourier-beam}) then becomes complex~\cite{Zwanziger:1973if,Ilderton:2017xbj}.
{Not all aspects of back-reaction can be captured by a \emph{single} final coherent state~\cite{Ekman:2020vsc}}, {so we focus here on the simplest case of complete depletion, where $\beta_s=0$ -- that is, the initial coherent state is completely absorbed during scattering.} The complex field (\ref{Fourier-beam}) is naturally given by retaining only positive energy (i.e.~incoming) modes in the Fourier expansion of the corresponding \emph{real} field.
 
Now, any electromagnetic field strength $F_{\mu\nu}$ can be decomposed into self-dual and anti-self-dual parts; these are just the $\pm i$-eigenspaces of the duality operation $F_{\mu\nu}\to \frac{1}{2}\epsilon_{\mu\nu\rho\sigma} F^{\rho\sigma}$ (or $\mathbf{E}\to\mathbf{B}$, $\mathbf{B}\to -\mathbf{E}$). A complex field is therefore one for which these two parts are not related by complex conjugation. From this perspective, the simplest example of a complex electromagnetic field is one which is purely \emph{self-dual}: a field in which only the self-dual field strength is non-zero. Any such field automatically solves the vacuum Maxwell equations (as they become equivalent to the Bianchi identity), but self-dual field configurations are in fact a classical integrable system, and can be described by free analytic data~\cite{Ward:1977ta,Belavin:1978pa,Ward:1990vs,Mason:1991rf}. {What is more, this integrability \emph{extends to the equations of motion} for massless fields coupled to self-dual backgrounds: the wavefunctions of massless fields, charged with respect to a self-dual background, are also determined by free analytic data~\cite{Penrose:1969ae,Hitchin:1980hp,Eastwood:1981jy,Ward:1990vs}.}

These theorems (typically stated in the language of twistor theory~\cite{Penrose:1967wn,Penrose:1972ia}) make self-dual fields remarkably tractable backgrounds, encoding interesting information about limits of vacuum amplitudes~\cite{Adamo:2021hno} and enabling the determination of exact wavefunctions and even higher-loop and high-multiplicity scattering amplitudes in several examples in both gauge theory and gravity~\cite{Dunne:2001pp,Dunne:2002qf,Dunne:2002qg,Adamo:2020syc,Adamo:2020yzi,Adamo:2022mev,Bogna:2023bbd,Adamo:2023fbj,Adamo:2024xpc,Garner:2024tis,Dixon:2024mzh,Dixon:2024tsb,Bittleston:2024efo}. {Consequently, we look for a \emph{self-dual field} which models a totally depleting focussed beam.}

\paragraph{From plane waves to flying-focus.}
Our first goal is to construct a flying focus beam with self-dual, positive frequency part.  This will, following the above, represent a focussed beam completely depleted during some interaction (to be chosen later). We work throughout in lightfront coordinates $x^\LCpm =(x^0 \pm x^3)/\sqrt{2}$, $x^\LCperp=(x^1,x^2)$. 

It is easily checked that any function $\varphi$ of the form 
\be\label{ff-scalar}
   \varphi(x) = \frac{f(\sigma)}{1+i k x^\LCp} \;, 
   \qquad \sigma  :=  x^\LCm - \frac{ik}{2}\frac{x^\LCperp x^\LCperp}{1+i k x^\LCp} \;,
\ee
with $k>0$ constant, is a solution of the wave equation in vacuum. Taking $f(\sigma) \propto \exp(-i\omega \sigma)$, for frequency $\omega>0$ gives a scalar flying focus wave with Gaussian focussing, in which the focus moves in the opposite direction to the phase velocity {at the speed of light}. We can extend this to a solution of Maxwell's equations as follows.

Following~\cite{Tod:1982mmp,Berman:2018hwd}, define the `spin raising' operator (vector indices in order $+,-,\perp$)
\be\label{spin}
    {\hat O}_\mu = \frac{1}{\sqrt{2}} ( \partial_1 + i \partial_2, 0, \partial_\LCm, i \partial_\LCm)\;.
\ee
{Acting with (\ref{spin}) on (\ref{ff-scalar}) produces the class of field}
\be\label{ff-maxwell}
    \mathcal{A}_\mu = \frac{a(\sigma)}{1+i k x^\LCp}\Big(-i k \frac{\epsilon_\LCperp x^\LCperp}{1+i k x^\LCp}, 0, \epsilon_\LCperp\Big) \;,
\ee
in which $a(\sigma)$ is arbitrary and $\epsilon_\LCperp=(1,i)/\sqrt{2}$. This is, as can be checked directly, a self-dual solution of the vacuum Maxwell equations. For the particular choice
\be\label{ff-beam}
a(\sigma)=\frac{E_0}{\omega}\,\e^{-i\omega\sigma}\,,
\ee
$\mathcal{A}_\mu$ is the positive frequency self-dual part of a circularly polarised flying focus beam, frequency $\omega$, focal width $w_0 =1/\sqrt{\omega k}$ and peak electric field~$E_0/2$, {where the focal spot counter-propagates at the speed of light with respect to the phase velocity.  Thus, $\mathcal{A}_{\mu}$ with $a(\sigma)$ given by \eqref{ff-beam} represents, as desired, a totally depleted} flying focus beam. For now, we will leave $a(\sigma)$ arbitrary, in which case the more general solution~\eqref{ff-maxwell} has been called a `focus wave'~\cite{Hillion92}.
{There are many other flying-focus solutions of Maxwell's equations, see e.g.~\cite{Ramsey2023,Formanek:2023zia} for beams with orbital angular momentum and different focal velocities.}

{Let us develop some intuition for both the fields we consider and the operator (\ref{spin}).} Returning to the Fourier representation of an arbitrary vacuum solution (\ref{Fourier-beam}), we work from here on in lightfront gauge $\varepsilon^s_\LCm(\ell)=0$, for which $\varepsilon^s_\mu = \big((\ell_1 + i s \ell_2)/\ell_\LCm,0,1,i s\big)/\sqrt{2}$.
{This expressions makes it immediately clear that the effect of the spin-raising operator on a scalar field is simply to insert, under its Fourier integral, a photon polarisation vector in lightfront gauge. In this way it lifts solutions of the wave equation to solutions of Maxwell's equations.}

{(Acting again with (\ref{spin}) on (\ref{ff-maxwell}) produces a self-dual \emph{metric} perturbation solving the vacuum Einstein equations; this allows us to extend our results to scattering in gravity, as will appear elsewhere~\cite{AdamoIldertonToAppear}, and provides an intriguing link between gravity and strong field QED.)}

{Turning now to the fields, we again consider (\ref{Fourier-beam}). A} real (fixed) plane wave is given by the choice of Fourier coefficients $\beta_s(\ell) = \alpha_s(\ell)\sim \delta^2(\ell_\LCperp) \ell_\LCm {a}_s(\ell_\LCm)$ for some temporal wave profile $a_s$. The focussed fields (\ref{ff-maxwell}) are given simply by (i)~restricting to only positive helicity modes, $s=1$, and (ii)~broadening the transverse delta functions of the plane wave Fourier coefficients into Gaussians of finite width proportional to $k \ell_\LCm$~\cite{Sezginer1985}.
{Doing so, the $\ell_\LCperp$-integrals in (\ref{Fourier-beam}) become Gaussian, so can be performed, and the $\ell_\LCm$-integral becomes a simple Fourier transform}, yielding the real field $A_\mu = \mathcal{A}_\mu + \mathcal{A}^\star_\mu$ with $\mathcal{A}_\mu$ as in (\ref{ff-maxwell}).
The important point is just that {the flying focus beam (\ref{ff-beam}) we consider} is a deformation of a plane wave by the parameter $k$, so comparison with plane waves will be immediate upon taking the $k\to0$ limit.
This will considerably aid the physical interpretation of our results.

{Finally, we note that the form of the gauge field $\mathcal{A}_\mu$ and its electromagnetic field strength $\mathcal{F}_{\mu\nu}=\partial_\mu \mathcal{A}_\nu - \partial_\nu \mathcal{A}_\mu$ offer some hint that scattering amplitudes computed in this self-dual background may be remarkably simple: it is easy to show that $\mathcal{A}^2 = 0$ and $\mathcal{F}_{\mu\sigma}{\mathcal{F}^\sigma}_\nu = 0$, so there are few invariants which can be constructed from the field, and on which physical results could depend.}

\paragraph{Particle wavefunctions in flying-focus beams.}
%
{We turn now to the} construction of wavefunctions {in the flying focus model (\ref{ff-maxwell})--(\ref{ff-beam})} which {describe}
initial and final-state particles. We make two simplifications which allow us to focus on the key message and present compact results. First, we assume that the scattered particles are ultrarelativistic, so that their mass can be neglected, but we make no assumption on the relative strength of the flying focus background, so our calculations apply equally to strong and weak fields. Second, we work in scalar, rather than spinor, QED.

As such, the desired wavefunctions are solutions of the Klein-Gordon equation $D^2 \phi=0$, where the covariant derivative is $D_\mu = \partial_\mu - i e \mathcal{A}_\mu$. A famous example is the Volkov solution for (real) plane wave backgrounds~\cite{Wolkow:1935zz}. {In the case of the depleting flying focus, with $\mathcal{A}_{\mu}$ given by \eqref{ff-maxwell}, theorems (cf., \cite{Penrose:1969ae,Hitchin:1980hp,Eastwood:1981jy,Ward:1990vs}) tied to the self-duality of this field guarantee that exact, explicit solutions to this wave equation must exist.} {Indeed, we find that the wavefunction of} an incoming (scalar) electron, momentum $p_\mu$, is described by 
\be\label{in-solution}
    \phi(x) = \frac{\e^{-iS}}{1+i k x^\LCp} \;,
\ee
in which $S$ is, writing $\sfp\equiv \epsilon_\LCperp p_\LCperp$, 
\be\label{S-solution}
    S = \frac{p_\LCperp x^\LCperp + p_\LCp x^\LCp}{1+i k x^\LCp}
    + p_\LCm \sigma
    + \frac{e \sfp}{p_\LCp} \int\limits^{\sigma}\!\ud s\, a(s) \;.
\ee
This wavefunction is remarkably simple, considering that it describes a particle interacting with, and absorbing energy from, a focussed field. The Volkov solution for a (depleting) plane wave is recovered as $k\to 0$.  The wavefunction for an outgoing electron, momentum $q_\mu$, is obtained as usual by solving the conjugate equation ${\bar D}^2\bar\phi = 0 $ where ${\bar D}_\mu = \partial_\mu + i e \mathcal{A}_\mu$ (note the background is not conjugated). The solution is given simply by sending $p_\mu \to - q_\mu$ in (\ref{S-solution}).

These wavefunctions have familiar properties: for instance, it is easily checked that $S$ is a Hamilton-Jacobi action obeying $(\partial S  + e\mathcal{A})^2 = 0$, and that the current
\be\label{eq-j}
    j_\mu(x) = {\bar \phi} D_\mu \phi - \phi {\bar D}_\mu{\bar \phi} \;,
\ee
is conserved as expected, $\partial_\mu j^\mu(x) =0$. 

\paragraph{Nonlinear Compton Scattering.}
{Our next goal is to show that our wavefunctions can be used in \emph{practical calculations} and yield clear, accessible results. We will demonstrate this via the simplest nontrivial example, namely the calculation of} the amplitude for
nonlinear Compton scattering, that is, photon emission from a charge interacting with an intense laser, see~\cite{E144:1996enr} and \cite{Abramowicz:2021zja}
 for past and future experiments, respectively.
 {(Clearly, it is unlikely that an entire beam will be absorbed by a single electron emitting a single photon. We present this calculation as a  proof-of-principle -- more phenomenologically relevant calculations will be presented elsewhere.)}

We consider an electron scattering from momentum $p_\mu$ to $q_\mu$ {on the beam (\ref{ff-maxwell})--(\ref{ff-beam}), totally depleting it}, and emitting a photon of momentum $\ell_\mu$ and helicity $s$. The Feynman rules for scalar QED yield the corresponding scattering amplitude as
\be\label{NLC-amp}
    S_{fi} = e \int\!\ud^4 x\, 
    \mathrm{e}^{i\ell\cdot x}
    {\bar \varepsilon}^\mu_s(\ell)
    j_\mu(x) \;,
\ee
in which $j_\mu(x)$ is the conserved current (\ref{eq-j}).

Our goal is to perform the integrals in (\ref{NLC-amp}). The apparent additional difficulty that all quantities become complex points, happily, to its own resolution: we can exploit the possibility of deforming the integrals in the complex plane. We begin by displacing the $x^\LCm$ contour by the non-trivial part of $\sigma$, and integrating along the line
\be
    x^\LCm \to x^\LCm + \frac{ik}{2}\frac{x^\LCperp x^\LCperp}{1+i k x^\LCp}\,, \quad x^\LCm \in \mathbb{R}\,.
\ee
This is valid provided the new contour does not cross any singularities of $a(\sigma)$ -- for the flying focus beam \eqref{ff-beam}, we only need to avoid the pole at $x^\LCp=i/k$, and we will see shortly that this is easy to control.  
Due to the photon wavefunction in (\ref{NLC-amp}), this choice of contour
renders the $x^\LCperp$--integrals \emph{Gaussian}. We therefore proceed by rotating the $x^\LCperp$ contours in the complex plane, effectively changing variable to $y^\LCperp=x^\LCperp/(1+i k x^\LCp)$.
The transverse integrals then produce
    \be\label{delta-regulated}
    (2 \pi)^2\,
    \frac{\e^{-\frac{(q+\ell-p)_\LCperp^2}{2k\ell_\LCm}}}{2 k \ell_\LCm \pi} =: {\hat\delta}^2_{\text{reg.}}(q+\ell-p) \;.
    \ee
As the notation suggests, this is a regulated delta function which, in the plane wave limit $k\to 0$, becomes $(2\pi)^2\delta^2_\LCperp(q+\ell-p)$, recovering transverse momentum conservation. Thus we see that the same Gaussian spread which turns a plane wave into a flying focus also generates a Gaussian spread in scattered particle momenta.

Next, the $x^\LCp$ integral can also be performed exactly, by changing variables from $x^\LCp$ to $z=x^\LCp/(1+i k x^\LCp)$, and integrating along real $z$. The singularity at $i/k$ in the complex $x^\LCp$-plane then becomes the point at infinity in the complex $z$-plane, and the integral over $z$ gives ($2\pi$ times) a delta function setting
    \be
        q_\LCp \to q_\LCp^\star := p_\LCp - \frac{(p_\LCperp-q_\LCperp)^2}{2 \ell_\LCm} \;,
    \ee
for the outgoing electron momentum.

{At this stage only the $x^\LCm$ integral remains and, exactly as in a (real) plane wave, should only be expected to be tractable for certain profiles $a(\sigma)$. Nevertheless, we can still make a further general statement} -- independent of the form of $a(\sigma)$, the amplitude for emission of a \emph{negative} helicity photon is a boundary term, which vanishes upon regularisation~\cite{Boca:2009zz}.
This is physically reasonable, since the electron has absorbed only \emph{positive} helicity photons from the background.
Thus, only the amplitude for emission of a positive helicity photon is non-trivial; this is given by, writing $\hat{\delta}(x):=2\pi\delta(x)$,
\be\label{SFI2}
\begin{split}
    S_{fi} = 2ie\, \ell_\LCm
    &\hat{\delta}(q_\LCp - q_\LCp^\star)
    {\hat\delta}^2_{\text{reg.}}(q+\ell-p)
     \\
    &\times \bigg[
    \frac{ \mathsf{p}\, q_\LCp -\mathsf{q}\,p_\LCp}{(\mathsf{p}-\mathsf{q})^2}
    \bigg]
    \int\!\ud x^\LCm\, \mathrm{e}^{i\Phi(x^\LCm)} \;,
\end{split}
\ee
where all dependence on the wave profile is contained in
\be
    \Phi(x^\LCm) = (\ell +q -p)_\LCm x^\LCm + e\Big[ \frac{\mathsf{q}}{q_\LCp} - \frac{\mathsf{p}}{p_\LCp}\Big] 
    \int^{x^\LCm}\!\!\!\ud s\, a(s) \;.
\ee
{Details of the remaining integral in (\ref{SFI2}) depend on $a(\sigma)$, again as for real fields, where different techniques are used to analyse amplitudes in e.g.~monochromatic fields~\cite{Berestetskii:1982qgu} constant crossed fields~\cite{Ritus:1985vta}, pulsed plane waves~\cite{Boca:2009zz,Seipt:2016rtk} and wavetrains~\cite{Heinzl:2010vg}.} For the flying focus beam \eqref{ff-beam}, we perform the integral by simply expanding the field-dependent part of the exponential as a power series, upon which the integral immediately returns
\begin{multline}\label{SFI3}
	\int\!\ud x\, \mathrm{e}^{i\Phi(x)} =\\  \sum_{n=0}^\infty 
    \frac{\hat\delta ( (\ell+q-p)_\LCm-  n \omega)}{n!}
    \bigg(\frac{e E_0}{\omega^2}\Big[ \frac{\mathsf{p}}{p_\LCp} - \frac{\mathsf{q}}{q_\LCp}\Big] \bigg)^n \;.
\end{multline}
{This sum has the same structure as that found in a monochromatic plane wave (depleting or not);} as for that case, the nonlinear Compton amplitude splits into a sum of `harmonic' contributions~\cite{Berestetskii:1982qgu}, in which the lightfront momentum transfer is a multiple of the driving field frequency $\omega$.
{This structure persists in e.g.~the photon emission spectrum, where delta functions are removed by integrating out final electron states and the sum over $n$ converges, as proven in~\cite{Harvey:2009ry}.} (Considering a finite wave-train instead of a genuinely monochromatic field broadens the delta comb in (\ref{SFI3}) into a multi-slit interference pattern, see~\cite{Heinzl:2010vg}.) {The only real difference compared to the well-studied real plane wave case} is that there is no `mass-shift'~\cite{Kibble:1965zza} in our wavefunctions, amplitude, or harmonics, for either the depleting flying focus or depleting plane wave. This is physically reasonable -- the mass shift imprint in the {photon spectrum~\cite{Khrennikov,Sakai:2015mra}} is inherited from periodicity in the background~\cite{Kibble:1965zza,Harvey:2012ie}, but this periodicity must be lost as the background is depleted.

\paragraph{Harmonic spectra.}
%
The simplest physical observables to extract from the amplitude (\ref{SFI2})--(\ref{SFI3}) are the allowed emitted photon frequencies. These are determined by the support of the delta functions in (\ref{SFI2}) and (\ref{SFI3}), as is familiar from the plane wave case~\cite{Berestetskii:1982qgu}. We solve
\begin{figure}[t!!]
\includegraphics[width=0.9\columnwidth]{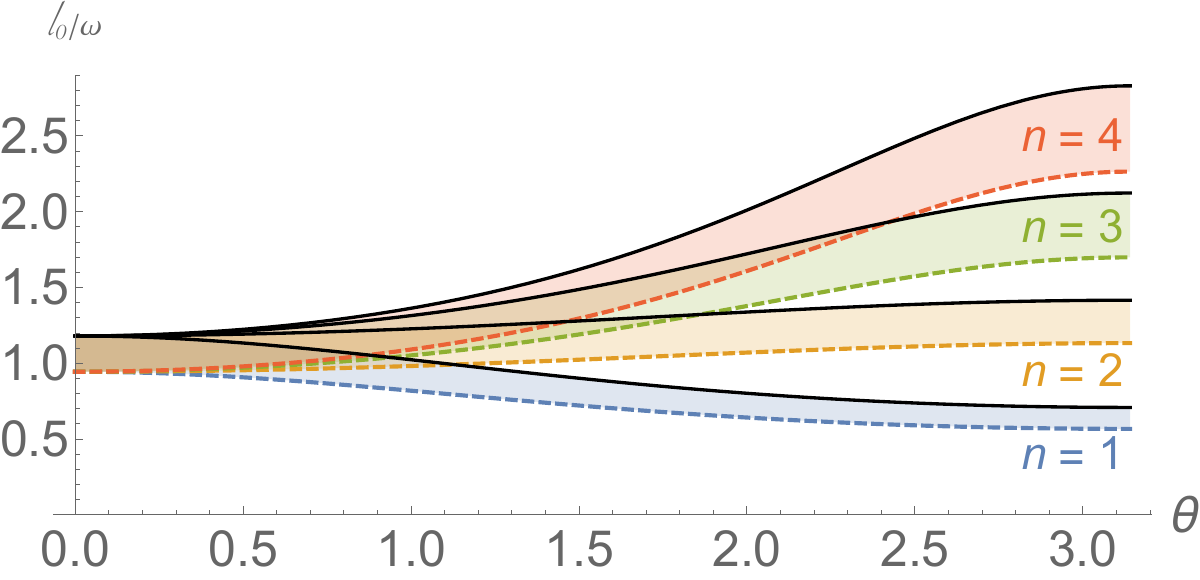}
\includegraphics[width=0.9\columnwidth]{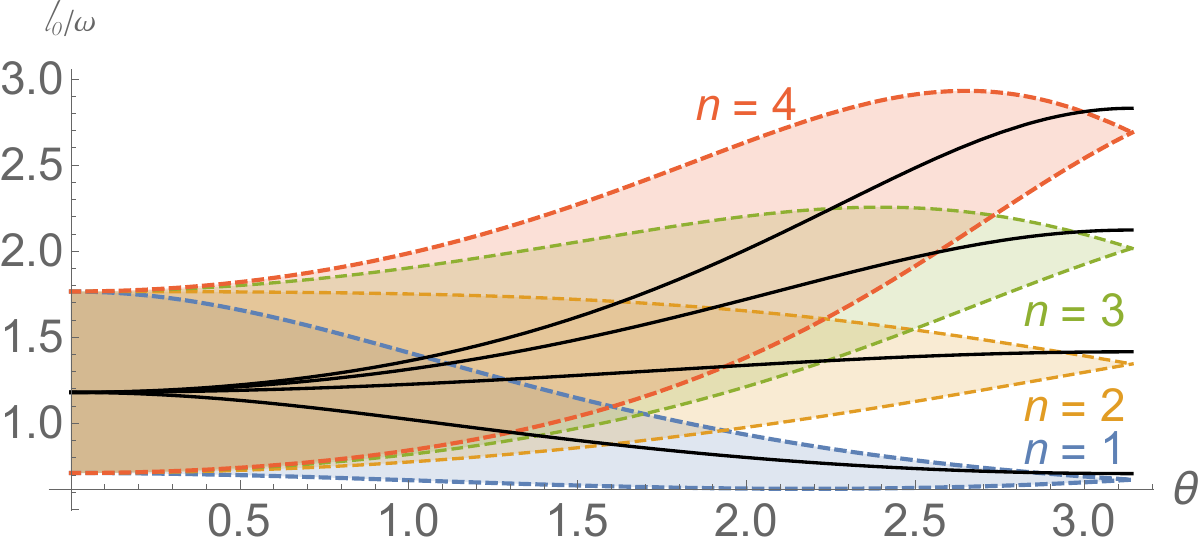}
    \caption{\label{fig:harmonics} Allowed photon frequencies $\ell_0$ as a function of emission angle $\theta$ and harmonic numbers ($n\in 1 \ldots 4$ to illustrate). 
    Solid black lines show the plane wave result, in which $\ell_0$ is uniquely determined by $\theta$ at each $n$. Coloured bands show the harmonic ranges in a flying focus beam, assuming the effective width determined by (\ref{delta-regulated}).  
    \emph{Top panel}: for electron momenta determined as below (\ref{harmonics-to-solve}) we see spectral broadening of the harmonics, with dashed lines indicating the effective cutoff~(\ref{FF-upper-range}). Here $p_\LCp/\omega = 5/3$ and $k/\omega = 1/3$. \emph{Bottom panel}:  for kinematics as in (\ref{parallel-case}) we see both spectral broadening around the plane wave limit for $\theta\simeq 0$ and, at $\theta\simeq \pi$, a distinct shift away from the plane wave results, the size of which increases with $n$ and $k$. Here $k/\omega = 1/12$, other parameters as above.}
\end{figure}
\be\label{harmonics-to-solve}
    \ell_\LCm + q_\LCm - p_\LCm  = n \omega \;,
\ee
for integer $n$, with $q$ evaluated on the support of the delta functions, taking $p_\LCperp = p_\LCm=0$ but $p_\LCp\not=0$, corresponding to a head-on collision for simplicity. In the plane wave limit, momentum conservation fixes $q_\LCperp = - \ell_\LCperp$ and $q_\LCp = p_\LCp - \ell_\LCp$ (with or without depletion). Converting back to Cartesian coordinates for clarity, (\ref{harmonics-to-solve}) then determines the emitted photon frequency $\ell_0$ as a function of the scattering angle $\theta$: 
\be\label{w-out-plane-wave}
    \ell_0 \to \ell_{\text{pw}} = \frac{n\sqrt{2}\omega}{1-\cos\theta +\frac{n\omega}{p_{\LCp}}(1+\cos\theta)} \;,
\ee
as is well known. 

In a flying focus beam, on the other hand, the lack of transverse momentum conservation, (\ref{delta-regulated}), means that the emitted photon frequency in each harmonic is \emph{not} uniquely determined by scattering angle. We therefore consider, in the spirit of recent experiments~\cite{Cole:2017zca,Poder:2017dpw,Los:2024ysw}, observations of both particles and emitted photons, scattered in particular directions. Again for a head-on collision, we select electron momenta of the form $q_\LCperp = -\ell_\LCperp + r_\LCperp$ with $r_\LCperp \ell_\LCperp=0$; that is, electrons scattered into kinematic regimes impossible in the plane wave case. The exponential suppression factor (\ref{delta-regulated}) (which is what replaces the transverse delta functions) effectively imposes the restriction $r_\LCperp^2 \lesssim 2 k \ell_\LCm$. Taking this as an upper bound and solving (\ref{harmonics-to-solve}), we find that the emitted frequencies lie between $\ell_{\text{pw}}$ and  
\be\label{FF-upper-range}
\bigg(1 - \frac{k}{p_\LCp}\bigg)\ell_{\text{pw}} \;.
\ee
This is illustrated in Fig.~\ref{fig:harmonics}, upper panel.

Other kinematics and observation angles allow richer structures.  For example, we could also look for photons scattered in the same transverse directions as in the plane wave case, but with other energies, such that
\be\label{parallel-case}
	q_\LCperp \in \left[ - \ell_\LCperp\bigg( 1 - \sqrt{\frac{k}{\ell_\LCp}}\bigg),
    - \ell_\LCperp\bigg( 1 + \sqrt{\frac{k}{\ell_\LCp}}\bigg)\right] \;,
\ee
in which the range is again effectively imposed by (\ref{delta-regulated}). In this case (\ref{harmonics-to-solve}) reduces to a quadratic equation, the solutions to which are plotted in Fig.~\ref{fig:harmonics}, lower panel.

\paragraph{Conclusions.}
{We have shown that modelling beam depletion via self-dual backgrounds can open the door to the inclusion of focussing effects in strong field QED.} We demonstrated this explicitly for {the flying focus beam (\ref{ff-maxwell})--(\ref{ff-beam}) and complete depletion},
providing exact particle wavefunctions in the depleting beam, despite the fact that the analogue of Volkov solutions for the undepleted background are not known.  {These are the only known exact solutions of an equation of motion in \emph{any} type of flying focus background.}  We used our wavefunctions to obtain the nonlinear Compton amplitude in the depleting flying focus beam, and extracted physical information on the emitted photon spectrum. {Despite having a complex background, the wavefunctions and amplitude are easy to analyse and have sensible properties.}
{Our results thus provide a concrete example of the potentially broad application of self-dual backgrounds to the modelling of \emph{real} physical effects.}

Turning to future work, we observe that the complete-depletion amplitude for scattering without emission vanishes in our approach, as is evident from its Feynman diagram expansion: momentum conservation forbids an electron from only absorbing photons, it must also emit. Again, this is physically sensible and \emph{independent} of loop order, suggesting that the inclusion of depletion may have implications for the Ritus-Narozhnyi conjecture~\cite{Ritus1,Narozhnyi:1980dc,Fedotov:2016afw,Mironov:2021ohk}.
We will explore this elsewhere. It would also be good to relax the idealisations we made in order to demonstrate the essential ideas here, e.g.~neglecting spin, and to explore other beam models. {As a final example, it is easy to check that $A_\mu =f(x^\LCm)(0,0,x^1+ix^2 ,ix^1-x^2)$, for $f$ an arbitrary complex function, is the self-dual version of an electromagnetic vortex~\cite{Bialynicki-Birula:2004bvr}. Scattering in the vortex is not necessarily phenomenologically relevant, but it would be intriguing to study nevertheless, due both to its solvability and its close links to gravity~\cite{Ilderton:2018lsf}.}

\begin{acknowledgments}
\textit{A.I.~thanks B.~King for useful discussions. The authors are supported by the STFC consolidated grant ``Particle Theory at the Higgs Centre" ST/X000494/1 (TA, AI), a Royal Society University Research Fellowship (TA), the Simons Collaboration on Celestial Holography MPS-CH-00001550-11 (TA), and the ERC Consolidator/UKRI Frontier grant ``TwistorQFT'' EP/Z000157/1 (TA).}
\end{acknowledgments}
\bibliography{FF-bib-2}
\end{document}